\documentclass[12pt]{iopart}

\begin{document}

\title[Reply to Comment]{Reply to Comment on ``Exact analytic solution for the generalized Lyapunov exponent
of the 2-dimensional Anderson localization''}

\author{V N Kuzovkov\dag, V Kashcheyevs\dag, W~von Niessen\ddag}

\address{\dag \ Institute of Solid State Physics, University of
Latvia, 8 Kengaraga Street, LV -- 1063 RIGA, Latvia}

\address{\ddag\ Institut f\"ur Physikalische und Theoretische Chemie,
Technische Universit\"at Braunschweig, Hans-Sommer-Stra{\ss}e 10,
38106 Braunschweig, Germany}

\ead{kuzovkov@latnet.lv}

\date{Received \today}
\begin{abstract}
We reply to  comments by P.Marko$\breve{s}$, L.Schweitzer and
M.Weyrauch [preceding paper] on our recent paper [J. Phys.:
Condens. Matter  {\bf 63}, 13777 (2002)]. We demonstrate that our
quite different viewpoints stem for the different physical
assumptions made prior to the choice of the mathematical formalism.
The authors of the Comment expect \emph{a priori} to see a single
thermodynamic phase while our approach is capable of detecting
co-existence of distinct pure phases. The limitations  of the
transfer matrix techniques for the multi-dimensional Anderson
localization problem are discussed.
\end{abstract}

\submitto{\JPCM} 

\maketitle

In our original publication \cite{Kuzovkov}  a conceptually new
view on the two-dimensional Anderson localization has been put
forward. The main message of our work can be summarized in one
sentence: the metal-insulator transition (MIT) `should be interpreted as a first-order phase
transition'\cite{Kuzovkov}. We believe that a misunderstanding of this
crucial point (which indeed `contradicts standard
wisdom'\cite{Comment}) has lead to the objections of P. Marko\v{s}
\emph{et al.}\cite{Comment}. In this Reply we elaborate on this
conceptual difference and then comment briefly on specific points
of the criticism mentioned in \cite{Comment}.

One of the cornerstones of the famous scaling theory of
localization \cite{Abrahams} and the subsequent numerical studies
is that for each particular energy and disorder strength the
system must be \emph{only in one} of the two phases: with either
localized, or extended wave functions. The main conclusion of our
work does not fit this phenomenological framework: we argue not for
the existence of a single extended phase, but for the
\emph{co-existence} of extended and localized phases for the same
energy and disorder strength. \emph{Which} of these two pure
phases is realized, depends on particular realization of the
disorder.

How does this conceptual difference translate into the difference
between our results and the results of conventional transfer
matrix calculations \cite{Comment,Pendry82}? Both approaches start
with the formal statistical ensemble which includes all
realizations of the disorder and thus describes both pure phases,
extended and localized. Averaging a physical quantity over this
ensemble produces an average between the properties of two pure
phases which is not capable of characterizing each phase
separately. This is the source of failure for the standard
approach: it considers such heterogeneous averages only and views
them as characterizing a single pseudo-pure phase. In such a
mixture, Lyapunov exponents of the pure insulating phase always
dominate\cite{Kuzovkov}, and this gives a false impression that
only localized solutions are present. This is how P. Marko\v{s}
\emph{et al.} pose the question: whether the average transfer
matrix $T^{(2)}$ describes the extended or the insulating phase?
Subsequently, the answer they find is not satisfactory: the
assumed pseudo-pure phase appears to be localized in 2D (as well
as in 3D) for any degree of the disorder.

Distinctive features of our mathematical  formalism (signal theory)
follow naturally from the requirement that \emph{the theory must be able to describe adequately
multiple pure phases} (in case if more than one such phase exist).
We have repeatedly emphasized this point in our original work by introducing
the term `multiplicity of solutions' (MS) \cite{Kuzovkov}.
The original problem is linear and the MS requirement
for an exact solution  puts severe restrictions on the form
of the mathematical formalism.
Two points are essential here: (i) a precise definition of the phase
in analytically exact theories is possible only in the thermodynamic limit\cite{Baxter}.
That is why our approach assumes an infinite system in the lateral direction ($M=\infty$ in terms of \cite{Comment})
from the very beginning.
(ii) We go beyond the description of heterogeneous averages $\langle \psi_{nm} \psi_{n'm'}
\rangle$
(signals in our terminology) by considering the fundamental function --- the filter $H(z)$.
In the region of phase co-existence the filter shows MS, corresponding to multiple subsets of
propagating signals. This is the point where one can start analyzing the properties
of distinct pure phases and not the mixture of them.

The transfer matrix approach \cite{Comment,Pendry82,MacKinnon81}
does not posses these crucial features. It starts with a quasi-1D
system and approaches the thermodynamic limit only asymptotically.
We have already emphasized  in our work (see Sec.~4.2 in
\cite{Kuzovkov}) that the quasi-1D transfer matrices loose the MS
property which is the key for a correct description of multiple
phases. Indeed, P.~Marko{\v{s}} \emph{et al.} note \cite{Comment}
that the thermodynamic limit is problematic in their approach:
`the limit $M\to \infty$ of the discrete model discussed here
bears various conceptual and technical difficulties.'
 Their suggested solution
(studying this limit numerically) is hardly adequate to overcome these difficulties, while
our approach treats the system as a truly multi-dimensional one from the starting point.

It is not a coincidence that we were able to solve the 1D case
both by the method of transfer matrix and the signal theory, while
for 2D the former method was abandoned \cite{Kuzovkov}. In 1D,
there is no phase transition and  the MS property plays no role,
while in higher dimensions the two methods are \emph{not}
equivalent.

Even within the transfer matrix approach, which should give the
properties of a single pseudo-phase, we do not fully agree with
the authors of the Comment \cite{Comment}. In the transfer matrix approach
(as well as in the signal theory), the fundamental Lyapunov
exponent is determined by the \emph{maximal} eigenvalue of the
transfer matrix (respectively, the \emph{maximal} root of the
equation for filter poles). Therefore the analysis of the
eigenvalues close to the unit circle\cite{Comment}  is not
sufficient to determine the phase diagram of the system.

We can also not agree with the statement \cite{Comment} concerning
the ambiguity of our averaging procedure. The latter has been described at
length in Sec.~3.2 of our original work\cite{Kuzovkov}, assuming
familiarity of the reader with only basic aspects of the signal theory and
linear algebra.

Regarding the last critical remark in \cite{Comment} we note
that since our approach does not seek to `detect a metallic phase'
but rather
to find the multiplicity  and the properties of pure phases,
there is no \emph{physical reason}
to prefer the analysis of the $\langle \ln  |\psi| \rangle$ over the analysis of the
second moment $\langle |\psi|^2 \rangle$.

Finally, we note that we are not alone in challenging the
prevailing view that there is no MIT transition for  2D Anderson
hamiltonian. The support from finite-size scaling studies  has
been put under the question mark \cite{Kantelhardt,Queiroz}. An even
more important challenge comes from experiment. A MIT has been
observed experimentally in 2D samples \cite{Abrahams2}, causing
significant new research activity. Direct electrostatic
probing \cite{Ilani1} and photoluminescence spectroscopy
\cite{Shashkin94} show a co-existence of localized and metallic
regions associated with 2D MIT, and new theories are put forward
to address this issue\cite{Meir00, Spivak03}. Ref.~\cite{Spivak03}
associates the phase separation with a \emph{first-order} phase
transition between a Fermi liquid and a Wigner crystal. At the
same time there is a growing evidence that the transition is
\emph{disorder-driven} and does not stem from electron-electron
interactions \cite{Meir00}. Thus there is a clear need for a revision
of the canonical point of view on the localization problem. This
revision should touch not only the numerical studies of a
tight-binding hamiltonian \cite{Kantelhardt,Queiroz}, but mainly
the scaling theory of localization which stands on
phenomenological grounds. Our exact analytic results for a
microscopic model show that one does not need to go beyond the
original framework of the Anderson hamiltonian in order to
describe a MIT and phase co-existence in 2D.



\section*{References}
\begin{thebibliography}{10}

\bibitem{Kuzovkov} V.N. Kuzovkov, W. von Niessen, V. Kashcheyevs and O.~Hein, J. Phys.:
Condens. Matter,  {\bf 14},  13777  (2002).

\bibitem{Comment} P. Marko\v{s}, L.Schweitzer and M.Weyrauch,
Comment.

\bibitem{Abrahams}
E. Abrahams, P.W. Anderson, D.C. Licciardello, and T.V.
Ramakrishnan, Phys. Rev. Lett. {\bf 42},  673  (1979).

\bibitem{Pendry82} J. B. Pendry, J. Phys. C: Solid State Phys., \textbf{15}, 3493 (1982).

\bibitem{Baxter} R.J.Baxter, Exactly Solved Models in Statistical
Mechanics (Academic Press, London, New York, 1982).

\bibitem{MacKinnon81} A. MacKinnon and B. Kramer, Phys.Rev. Lett.,
{\bf 21}, 1546 (1981).

\bibitem{Kantelhardt} J. W. Kantelhardt and A. Bunde, Phys. Rev. B \textbf{66}, 035118 (2002).

\bibitem{Queiroz} S. L. A. {de Queiroz}, Phys. Rev. B \textbf{66}, 195113  (2002).


\bibitem{Abrahams2}
E. Abrahams, S.V. Kravchenko, M.P. Sarachik, Rev. Mod. Phys. {\bf
73}, 251 (2001) and references there-in.

\bibitem{Ilani1} S. Ilani, A. Yacoby, D. Mahalu and H. Shtrikman, Science {\bf 292}, 1354 (2001).


\bibitem{Shashkin94} A.A. Shashkin, V.T.~Dollgopolov, G.V.~Kravchenko,
M.~Wendel, R.~Schuster, J.P.~Kotthaus, P.J.~Haug, K.~von Klitzing,
K.~Ploog, N.~Nickel, and W.~Schlapp, Phys. Rev. Lett. \textbf{73},
3141 (1994).


\bibitem{Meir00} Y. Meir, Phys. Rev. B \textbf{61}, 16470 (2000)
 and references therein.

\bibitem{Spivak03} B. Spivak, Phys. Rev. B \textbf{67}, 125205 (2003).

\end{thebibliography}
\end{document}